\documentclass[aps,prl,groupedaddress,showpacs]{revtex4}
\usepackage{graphicx,epsf,amssymb}
\newcommand{\be}{\begin{equation}} 
\newcommand{\ee}{\end{equation}}

\newcommand{\bea}{\begin{eqnarray}} 
\newcommand{\eea}{\end{eqnarray}}

\newcommand{\NeqFour}{{\cal N} =4}
\newcommand{\F}{{\cal{F}}}

\newif\ifdraft
\drafttrue
\newif\ifpreprint
\preprinttrue

\def\NeqFour{{\cal N}=4}
\def\NeqOne{{\cal N}=1}

\def\nf{n_{\mskip-2mu f}}
\def\Nc{N_{c}}

\def\sandp#1.#2.#3{%
\left\langle\smash{#1}{\vphantom1}^{-}\right|{#2}%
\left|\smash{#3}{\vphantom1}^{+}\right\rangle}
\def\sandpp#1.#2.#3{%
\left\langle\smash{#1}{\vphantom1}^{+}\right|{#2}%
\left|\smash{#3}{\vphantom1}^{+}\right\rangle}
\def\sandmm#1.#2.#3{%
\left\langle\smash{#1}{\vphantom1}^{-}\right|{#2}%
\left|\smash{#3}{\vphantom1}^{-}\right\rangle}
\def\spab#1.#2.#3{\sandmm#1.#2.#3}
\def\spba#1.#2.#3{\sandpp#1.#2.#3}
\def\spaa#1.#2.#3.#4{\sandmp#1.{#2#3}.#4}
\def\spbb#1.#2.#3.#4{\sandpm#1.{#2#3}.#4}
\def\spa#1.#2{\left\langle#1\,#2\right\rangle}
\def\spb#1.#2{\left[#1\,#2\right]}
\def\spash#1.#2{\vphantom{\hat K}\spa{\smash{#1}}.{\smash{#2}}}
\def\spbsh#1.#2{\vphantom{\hat K}\spb{\smash{#1}}.{\smash{#2}}}
\def\lor#1.#2{\left(#1\,#2\right)}
\def\sand#1.#2.#3{%
\left\langle\smash{#1}{\vphantom1}^{-}\right|{#2}%
\left|\smash{#3}{\vphantom1}^{-}\right\rangle}
\def\sandpp#1.#2.#3{%
\left\langle\smash{#1}{\vphantom1}^{+}\right|{#2}%
\left|\smash{#3}{\vphantom1}^{+}\right\rangle}
\def\sandpm#1.#2.#3{%
\left\langle\smash{#1}{\vphantom1}^{+}\right|{#2}%
\left|\smash{#3}{\vphantom1}^{-}\right\rangle}
\def\sandmp#1.#2.#3{%
\left\langle\smash{#1}{\vphantom1}^{-}\right|{#2}%
\left|\smash{#3}{\vphantom1}^{+}\right\rangle}

\def\tr{\mathop{\rm tr}\nolimits}

\newbox\SlashedBox 
\def\slashed#1{\setbox\SlashedBox=\hbox{#1}
\hbox to 0pt{\hbox to 1\wd\SlashedBox{\hfil/\hfil}\hss}#1}
\def\hboxtosizeof#1#2{\setbox\SlashedBox=\hbox{#1}
\hbox to 1\wd\SlashedBox{#2}}

\newbox\charbox
\newbox\slabox
\def\s#1{{      
        \setbox\charbox=\hbox{$#1$}
        \setbox\slabox=\hbox{$/$}
        \dimen\charbox=\ht\slabox
        \advance\dimen\charbox by -\dp\slabox
        \advance\dimen\charbox by -\ht\charbox
        \advance\dimen\charbox by \dp\charbox
        \divide\dimen\charbox by 2
        \raise-\dimen\charbox\hbox to \wd\charbox{\hss/\hss}
        \llap{$#1$}
}}

\def\eqn#1{eq.~(\ref{#1})}

\def\e{\epsilon}

\def\F#1#2{\,{{\vphantom{F}}_{#1}F_{#2}}}

\def\tree{{\rm tree}}
\def\oneloop{{1 \mbox{-} \rm loop}}

\def\cg{c_\Gamma}

\def\sandp#1.#2.#3{%
\left\langle\smash{#1}{\vphantom1}^{+}\right|{#2}%
\left|\smash{#3}{\vphantom1}^{+}\right\rangle}
\def\ksl{\s{k}}
\def\Ksl{\s{K}}

\def\V{V}
\def\F{F}

\def\Remaining{{\hat {R}}}

\newbox\ourfigbox
\def\SizedFigureWithCaption#1#2#3{%
\setbox\ourfigbox
  \hbox{\hss\epsfxsize #1 \epsfbox{#2}\hss}
\hbox to \wd\ourfigbox{\vbox{\noindent\copy\ourfigbox\break
\vskip -6mm      \hbox to \wd\ourfigbox{\hss#3\hss}}}
}

\def\llongrightarrow{%
\relbar\mskip-0.5mu\joinrel\mskip-0.5mu\relbar
     \mskip-0.5mu\joinrel\longrightarrow}
\def\inlimit^#1{\buildrel#1\over\llongrightarrow}
\def\dash{\hbox{-\kern-.02em}}

\begin{document}
\hfuzz 10 pt


\ifpreprint
\noindent
Saclay/SPhT--T05/149
\hfill 
\fi

\title{All-Multiplicity One-Loop Corrections to MHV Amplitudes in QCD}

\author{Darren Forde} 
\affiliation{Service de Physique Th\'eorique,
   CEA--Saclay\\ 
          F--91191 Gif-sur-Yvette cedex, France}

\author{David A. Kosower} 
\affiliation{Service de Physique Th\'eorique,
   CEA--Saclay\\ 
          F--91191 Gif-sur-Yvette cedex, France}

\date{September 2005}

\begin{abstract}
We compute the complete one-loop corrections to the simplest 
class of QCD gluon amplitudes, those with two color-adjacent
opposite-helicity
external particles.  We present results for an arbitrary number of
external legs.  This is the first all-$n$ computation of amplitudes that
enter into collider observables at next-to-leading order.
The computation uses the recently-developed 
on-shell recursion
relations for the rational parts, along with older unitarity-based
results for the cut-containing terms.
\end{abstract}

\pacs{11.15.Bt, 11.55.Bq, 12.38.Bx \hspace{1cm}}

\maketitle



\renewcommand{\thefootnote}{\arabic{footnote}}
\setcounter{footnote}{0}


The upcoming experimental program at CERN's Large Hadron Collider will
require many new calculations of one-loop corrections to QCD
and QCD-associated processes such as $W+\hbox{\rm\ multi-jet}$ production. 
 The development of new tools for
such computations is thus an important topic for theorists.

Recently, Bern, Dixon, and one of the authors presented 
a new approach~\cite{MMPaper} to 
computing complete one-loop scattering amplitudes in 
non-supersymmetric gauge theories such as
QCD.  The approach makes use of on-shell recursion relations to
systematize a unitarity-factorization bootstrap earlier applied
to the computation of the
one-loop scattering amplitudes needed for $Z \rightarrow 4$ jets and
$p p \rightarrow W + 2$ jets at next-to-leading order (NLO) in the QCD
coupling~\cite{ZFourPartons}.  In the combined approach, the cut-containing
terms (logarithms and polylogarithms) are computed using the 
unitarity-based method~\cite{Neq4Oneloop,Neq1Oneloop,MassiveUnitarity,%
UnitarityMethod,BBSTQCD}, with four-dimensional tree-level
amplitudes as input.  The remaining rational-function pieces 
are then computed via a factorization bootstrap, here
in the form of an on-shell recursion
relation~\cite{BCFRecursion,BCFW,OnShellRecursionI,QPaper,MMPaper}.  
(The use
of recursion relations supersedes the use of ans\"atze~\cite{ZFourPartons} 
for constructing the purely rational terms.)

The unitarity-based method has proven to be a powerful method for
computing the logarithmic and polylogarithmic terms in gauge theory
amplitudes at one and two loops.  Indeed, in massless supersymmetric theories
the complete one-loop amplitudes may be determined from the
four-dimensional cuts~\cite{Neq1Oneloop}.  The method has been used
to produce a variety of all-multiplicity 
results.
One can also use the unitarity-based
method to determine complete amplitudes, including
all rational pieces~\cite{MassiveUnitarity,SelfDualYM,DDimUnitarity,BMST} by
applying full $D$-dimensional unitarity, within dimensional regularization~\cite{HV}
($D = 4-2\e$).  In order to do so at one loop, one must compute
tree amplitudes where two of the
momenta are in $D$ dimensions.  In the case at hand,
one-loop amplitudes containing
only external gluons, these tree amplitudes can be interpreted as
four-dimensional amplitudes with two massive scalar legs.   The 
analytic computation
of these scalar amplitudes (even with recent 
advances~\cite{BadgerMassive,AllMultiplicityMassiveScalar}) and
their use in cuts is more complicated than that of massless four-dimensional
amplitudes.

The on-shell recursion relations~\cite{BCFRecursion} have their origins in alternate
representations of tree amplitudes emerging from one-loop 
calculations~\cite{NMHV,BCF7,RSVNewTree}.
The proof in ref.~\cite{BCFW} made it clear, however, that their applicability
is much wider.  The only ingredients which are needed are complex analysis, along
with a knowledge of the factorization properties of amplitudes in {\it complex\/}
momenta.  These properties are more subtle at loop level than at tree
level~\cite{OnShellRecursionI,QPaper}, 
but this does not affect the amplitudes we compute here.

We follow the notation of ref.~\cite{MMPaper}, and compute
the color-ordered on-shell one-loop amplitude $A_n(1^-,2^-,3^+,\ldots,n^+)$,
that is the configuration with adjacent negative-helicity gluons.
The corresponding tree-level amplitudes were conjectured by 
Parke and Taylor~\cite{ParkeTaylor,MPX} and proven by 
Berends and Giele using off-shell 
recursion relations~\cite{BGRecursion,LCRecursion}.
As explained in ref.~\cite{Neq4Oneloop}, 
this amplitude (along with related amplitudes
with non-adjacent negative-helicity gluons) would suffice to obtain
the full color-dressed amplitude with two negative-helicity gluons.
The tree-level amplitude for this configuration does not vanish.  
Accordingly, the one-loop amplitude we compute contributes to physical
observables at NLO, through its interference with the tree-level
amplitude.  For the same reason,
the one-loop amplitude has both ultraviolet and infrared divergences,
for which we use dimensional regularization.  The color-ordered amplitude
can be written in terms of functions which have an interpretation in
terms of supersymmetric and non-supersymmetric contributions~\cite{Neq4Oneloop},
\begin{eqnarray}
A_{n;1}^{\rm QCD} &=& 
\cg \biggl[(\V_n^g+4\V_n^f+\V_n^s) A^\tree_n + i (4\F_n^f+\F_n^s) 
  - {\nf \over \Nc} \Bigl( A_n^\tree (\V_n^s + \V_n^f) + 
                                   i (\F_n^s + \F_n^f) \Bigr) \biggr] \,,
\label{QCDAmplitude}
\end{eqnarray}
where $\Nc$ is the number of colors and $\nf$ the number of quark
flavors.  All functions in this expression except for $\F^s$ have been
computed previously, in refs.~\cite{Neq4Oneloop,Neq1Oneloop}. 
The $X^g$ contributions
correspond to the $\NeqFour$ supersymmetric amplitude, and 
$-X^f$ to the $\NeqOne$ supersymmetric amplitude.  

The branch cut-containing contributions 
to $\F^s$ were also computed in ref.~\cite{Neq1Oneloop}, and serve as the
basis for the present computation of the purely-rational terms.  (The
corresponding contributions to amplitudes with non-adjacent negative
helicities were computed in ref.~\cite{BBSTQCD}.)  The computation proceeds
along the lines of ref.~\cite{MMPaper}.  The pure cut terms have
spurious singularities not present in the amplitude as a whole.  These
can be removed in the case at hand by replacing, for example,
\begin{equation}
{\ln(s_1/s_2)\over (s_2-s_1)^3} \longrightarrow {L_2(s_1/s_2)\over s_2^3}\,,
\end{equation} where
\begin{equation}
L_2(r) = {\ln r - (r-1/r)/2\over (1-r)^3}\,.
\end{equation}
This removal was in fact already 
carried out in the form given in ref.~\cite{Neq1Oneloop}.
To set up a recursion relation, one picks two spinors to shift; we take
these to be the two negative helicities $(1,2)$, so that
\begin{equation}
|1^-\rangle \rightarrow |1^-\rangle - z|2^-\rangle \,,\qquad
|2^+\rangle \rightarrow |2^+\rangle + z|1^+\rangle \,, 
\label{SpinorShift}
\end{equation}
and other spinors are unaffected.  We can verify that $\hat C_n(z)$, given
below in eq.~\ref{CHat}, vanishes as required when $z\rightarrow\infty$.
There are two separate rational contributions that we must compute:
the `direct-recursive' or `diagrammatic' terms, similar to the 
recursive terms in a tree-level
computation, but here using loop vertices in addition to tree ones; and
`overlap' contributions.  The latter contributions remove the double-counting
due to the presence of rational terms in $\hat C_n$, and are
computed by taking the residues of those rational terms.  

With our choice of shift,
the diagrammatic contributions themselves are of two types,
\begin{eqnarray}
&&\hspace*{-0.8cm}R_n(1^-,2^-,3^+,...,n^+) =
\nonumber\\
&&\sum_{j=4}^{n-1}\bigg[A^\tree_{n-j+2}(\hat{1}^-,\hat{K}_{2\cdots j}^-,(j+1)^+,...,n^+)
\frac{1}{K^2_{2\cdots j}}
  A^\oneloop_{j}(-\hat{K}_{2\cdots j}^+,\hat{2}^-,3^+,...,j^+)
\nonumber\\
&&\hphantom{ \sum_{j=3}^{n-1}! }
  +A^\oneloop_{n-j+2}(\hat{1}^-,\hat{K}_{2\cdots j}^+,(j+1)^+,...,n^+)
\frac{1}{K^2_{2\cdots j}}A^\tree_j(-\hat{K}_{2\cdots j}^-,\hat{2}^-,3^+,...,j^+)
\bigg]
\nonumber\\
&&\hphantom{=}
  +R_{n-1}(\hat{1}^-,\hat{K}_{23}^-,4^+,...,n^+)\frac{1}{K^2_{23}}
A^\tree_3(-\hat{K}_{23}^+,\hat{2}^-,3^+)\,,
\label{eq:full_rec_rel}
\end{eqnarray}
where the one-loop amplitudes correspond to the contributions of
internal scalars, and where
we have not written out terms that vanish because of the vanishing
of the $A_3^\tree(\hat 1^-,\hat{K}^+,n^+)$ vertex with our choice of shift.
The first type, the terms summed over $j$,
 consists of all contributions with multiparticle poles or
a $\spb1.n$ pole.  (The $j=3$ term drops
out because the internal-scalar contributions to 
$A^\oneloop(-\hat K_{23}^+,\hat 2^-,3^+)$ vanish.)
These
involve
vertices whose all-$n$ form is known, either tree-level MHV amplitudes,
or $A_n^{\oneloop}(1^-,2^+,\ldots,n^+)$ (whose form we take from 
ref.~\cite{QPaper} in preference to the original calculation~\cite{Mahlon}).  
The second type, the last term in \eqn{eq:full_rec_rel},
is the contribution in the $\spa2.3$ channel,
containing a three-point vertex, $A_3^\tree(-\hat K_{23}^+,\hat 2^-,3^+)$
multiplied by all rational terms from our desired amplitude --- 
$A_{n-1}^{\oneloop}(\hat 1^-,-\hat K^-,4^+,\ldots,n^+)|_{\rm rational}$ ---
with one {\it fewer\/} positive-helicity external legs.  
The structure of this term is the same
as one of the terms in the recursion for the all-multiplicity amplitude
with a massive scalar pair and an adjacent negative-helicity 
gluon~\cite{AllMultiplicityMassiveScalar}.  We can proceed
here using the same approach, repeatedly inserting the recursion
relation into itself, at each step reducing the number of positive-helicity
legs in the unknown amplitude.  We thereby `unwind'
the recursion, rewriting it in the form,
\begin{equation}
\sum_{j=1}^{n-3} \prod_{r=2}^j 
   {A_{3\,r}^\tree\over K_{2\cdots (j+1)}^2} \times T_{n-j+1}(\hat 1^-,\ldots,n^+),
\end{equation}
in which $T_j$, encompassing the first type of terms above,
 now contains only known functions --- tree amplitudes, loop
vertices, rational terms from $\hat C_j$, and terms from 
overlap contributions.

We can then write the result for the unrenormalized amplitude
$A^\oneloop_{n,s} = \cg(\V^s A^\tree_n+i\F^s)$ in the following form,
\begin{eqnarray}
\cg\bigl[\V_n^s A^\tree_n+i\F_n^s\bigr] 
= \cg \bigl[ \hat C_n + \Remaining_n \bigr] 
+ {1\over3} A_n^{\NeqOne{\rm \ chiral}} + {2\over9} A_n^\tree
\end{eqnarray}
where $\hat C_n$ are the cut-containing contributions computed in 
ref.~\cite{Neq1Oneloop},
completed so as to remove $s_1\rightarrow s_2$ spurious singularities,
\begin{eqnarray}
\displaystyle
&&\hspace*{-0.8cm}
  \hat C_n = -{1\over3 s_{12}^3} A^\tree(1^-,2^-,3^+,\ldots,n^+)
\nonumber\\
&&\times \sum_{m=4}^{n-1} 
  \frac{L_2\bigl( (-s_{2\cdots (m-1)})/(-s_{2\cdots m})\bigr)}{s_{2\cdots m}^3}
\tr_{+}\bigl[\ksl_1\ksl_2\ksl_m\Ksl_{m\cdots 1}\bigr]
      \tr_{+}\bigl[\ksl_1\ksl_2\Ksl_{m\cdots 1}\ksl_m\bigr]
\tr_{+}\bigl[\ksl_1\ksl_2(\ksl_m\Ksl_{m\cdots 1}-\Ksl_{m\cdots 1}\ksl_m)\bigr]
\label{CHat}
\end{eqnarray}

The computations then yield,
\def\spacerI{%
\hphantom{\times \sum_{i_1=1}^{n-4} \Bigg(\sum_{i_2=i_1+3}^{n-1} \biggl[\hskip -2mm])}}
\def\spacerII{%
\hphantom{\times \sum_{i_1=1}^{n-4} \Bigg()}}
\def\spacerIII{\spacerII\hskip 10mm}
\def\vspacer{\vphantom{\sum_j^i}}
\begin{eqnarray}
&&\hspace*{-0.8cm}\Remaining_{n}(1^-,2^-,3^+,\ldots,n^+)
=\frac{1}{3}A^\tree(1^-,2^-,3^+,\ldots,n^+)
\nonumber\\
&&\times\sum_{i_1=1}^{n-4}
 \Bigg(\sum_{i_2=i_1+3}^{n-1} 
           \biggl[
            C_1(n;i_1,i_2) \Bigl( T_{1}(n;i_1,i_2,i_2)+ T_{1}(n;i_1,i_2,i_2+1)\Bigr)
\nonumber\\ &&\spacerI
           {}+C_2(n;i_1,i_2) \Bigl( T_{2a}(n;i_1,i_2)+ T_{2b}(n;i_1,i_2)\Bigr)
\nonumber\\ &&\spacerI
           {}+C_3(n;i_1,i_2) \Bigl( T_{3a}(n;i_1,i_2)+ T_{3b}(n;i_1,i_2)
                                  + T_{3c}(n;i_1,i_2)\Bigr)
            \biggr] 
           {}+ T_4(n;i_1)
 \Bigg)
\label{Rhat}
\end{eqnarray}
In this equation, 
\begin{eqnarray}
C_1(n;i_1,i_2) &=& 
  \displaystyle
  \frac{\spa{(i_1+1)}.{(i_1+2)}}
     {\sandmp{1}.{\Ksl_{(i_2+1)\cdots n}\Ksl_{(i_1+2)\cdots i_2}}.{(i_1+1)}%
      \sandmp{1}.{\Ksl_{(i_2+1)\cdots n}\Ksl_{(i_1+3)\cdots i_2}}.{(i_1+2)}},
\nonumber\\
C_2(n;i_1,i_2) &=& 
  \displaystyle
  \frac{\spa{i_2}.{(i_2+1)} C_1(n;i_1,i_2)}
     {s_{(i_1+2)\cdots i_2}%
      \sandmp{1}.{\Ksl_{2\cdots (i_1+1)}\Ksl_{(i_1+2)\cdots i_2}}.{(i_2+1)}%
      \sandmp{1}.{\Ksl_{2\cdots (i_1+1)}\Ksl_{(i_1+2)\cdots (i_2-1)}}.{i_2}},
\label{Coefficients}\\
C_3(n;i_1,i_2) &=& 
  \displaystyle
  s_{(i_1+2)\cdots i_2}^4 C_2(n;i_1,i_2).
\nonumber
\end{eqnarray}

The terms $T_i$ are given by,
\begin{eqnarray}
&&\hspace*{-0.8cm}
T_{1}(n;i_1,i_2,j)=\\
&&
  \displaystyle
  \frac{s_{(i_1+2)\cdots i_2}\spa{1}.{j}%
     \sandmp{1}.{\Ksl_{(i_2+1)\cdots n}\Ksl_{(i_1+2)\cdots i_2}}.{j}%
  \sandmp{1}.{\Ksl_{2\cdots i_2}\Ksl_{(i_1+2)\cdots i_2}(\ksl_{j}\Ksl_{2\cdots (j-1)}-\Ksl_{(i_1+2)\cdots (j-1)}\ksl_{j})}.{1}
    }
  {2\sandmp{1}.{\Ksl_{2\cdots (i_1+1)}\Ksl_{(i_1+2)\cdots i_2}}.{j}^2};
\nonumber
\end{eqnarray}
(Note that $T_{1}(n;i_1,n-1,n)=0$.)

\begin{eqnarray}
&&\hspace*{-0.8cm} T_{2a}(n;i_1,i_2) =
  \sum_{l=(i_1+2)}^{i_2}
 \sandmp{1}.{\Ksl_{(i_2+1)\cdots n}\Ksl_{(i_1+2)\cdots i_2}}.{l} f_1(n;l,i_1,i_2);\\
%
&&\hspace*{-0.8cm} T_{2b}(n;i_1,i_2) =
-\sum_{l=(i_1+3)}^{i_2-1}\sum_{p=l+1}^{i_2}
 \frac{f_2(n;l,p;i_1,i_2)}
      {\sandmp{1}.{\Ksl_{(i_2+1)\cdots n}\Ksl_{(i_1+2)\cdots i_2}\Ksl_{(i_1+2)\cdots (l-1)}\Ksl_{l\cdots (p-1)}}.{p}}
\nonumber\\
 \nonumber\\
&&\spacerIII\times
  \frac{\spa{(l-1)}.{l}%
   \sandmp{1}.{\Ksl_{(i_2+1)\cdots n}\Ksl_{(i_1+2)\cdots i_2}\Ksl_{l\cdots p}\Ksl_{(i_1+2)\cdots p}\Ksl_{(i_1+2)\cdots i_2}\Ksl_{2\cdots i_2}}.{1}^3}
      {s_{l\cdots p}\sandmp{1}.{\Ksl_{(i_2+1)\cdots n}\Ksl_{(i_1+2)\cdots i_2}\Ksl_{(i_1+2)\cdots p}\Ksl_{l\cdots p}}.{(l-1)}}
\\
&&\spacerIII\times
 \frac{\sandmp{1}.{\Ksl_{2\cdots i_2}\Ksl_{(i_1+2)\cdots i_2}\Ksl_{(i_1+2)\cdots (l-1)}[\mathcal{F}(l,p)]^2\Ksl_{(i_1+2)\cdots p}\Ksl_{(i_1+2)\cdots i_2}\Ksl_{2\cdots i_2}}.{1}}
             {\sandmp{1}.{\Ksl_{(i_2+1)\cdots n}\Ksl_{(i_1+2)\cdots i_2}\Ksl_{(i_1+2)\cdots p}\Ksl_{(l+1)\cdots p}}.{l}};
\nonumber\\
&&\hspace*{-0.8cm} T_{3a}(n;i_1,i_2) =
 \displaystyle
 \sum_{l=i_2+1}^{n-1}
  \frac{\spa{1}.{l}\spa{1}.{(l+1)}
      \sandmp{1}.{\Ksl_{l{}(l+1)}\Ksl_{(l+1)\cdots n}}.{1}}{\spa{l}.{(l+1)}};\\
%
&&\hspace*{-0.8cm} T_{3b}(n;i_1,i_2) =
 \displaystyle
 \frac{\sand{1}.{\Ksl_{2\cdots i_2}}.{(i_2+1)}%
         \sandmp{1}.{\Ksl_{2\cdots (i_1+1)}\Ksl_{2\cdots i_2}}.{1}%
         \spa{1}.{(i_2+1)}^2}
        {\sandmp{1}.{\Ksl_{2\cdots (i_1+1)}\Ksl_{(i_1+2)\cdots i_2}}.{(i_2+1)}};
\\
%
&&\hspace*{-0.8cm} T_{3c}(n;i_1,i_2) =
\displaystyle
 \sum_{l=i_2+1}^{n-2}\sum_{p=l+1}^{n-1}
  \frac{\sandmp{1}.{\Ksl_{l\cdots p}\Ksl_{(p+1)\cdots n}}.{1}^3}
       {\sandmp{1}.{\Ksl_{(p+1)\cdots n}\Ksl_{(l+1)\cdots p}}.{l}}
\nonumber\\
&&\spacerIII\times\frac{\spa{p}.{(p+1)}%
 \sandmp{1}.{\Ksl_{2\cdots (l-1)}[\mathcal{F}(l,p)]^2\Ksl_{(p+1)\cdots n}}.{1}%
          f_3(n;l,p,i_1,i_2)}
         {s_{l\cdots p}\sandmp{1}.{\Ksl_{2\cdots (l-1)}\Ksl_{l\cdots (p-1)}}.{p}%
               \sandmp{1}.{\Ksl_{2\cdots (l-1)}\Ksl_{l\cdots p}}.{(p+1)}};\\
%
&&\hspace*{-0.8cm} T_{4}(n;i_1) = 
 \displaystyle
-\frac{\spb{(i_1+2)}.{(i_1+3)}\spa{(i_1+3)}.1}
   {2\sand{1}.{\Ksl_{2\cdots(i_1+1)}}.{(i_1+2)}}.
\end{eqnarray}

\newlength{\widest}
\settowidth{\widest}{\hbox{$\hphantom{\displaystyle\hskip 10mm\times%
       \sandmp{1}.{\Ksl_{2\cdots i_2}\Ksl_{(i_1+2)\cdots i_2}%
                   \Ksl_{l,l+1}\Ksl_{(i_1+2)\cdots l}\Ksl_{(i_1+2)\cdots i_2}%
                   \Ksl_{2\cdots i_2}}.{1} }$}}
The $f_i$ appearing in the above equations are given by,
\begin{eqnarray}
&&\hspace*{-0.8cm} f_1(n;l,i_1,i_2) =
\nonumber\\
&&
\left\{\begin{array}{ll}
 \displaystyle
  -s^2_{(i_1+2)\cdots i_2}%
  \sandmp{1}.{\Ksl_{(i_1+2)\cdots i_2}\Ksl_{2\cdots (i_1+1)}}.{1}\\
  \hskip 10mm
 \displaystyle
\vspacer%
  \times
  \frac{%
       \sandmp{1}.{\Ksl_{2\cdots i_2}\Ksl_{(i_1+2)\cdots (i_2-1)}}.{i_2}
       \sandpp{i_2}.{\Ksl_{2\cdots (i_2-1)}}.{1}}
      {\sandmp{1}.{\Ksl_{2\cdots (i_1+1)}\Ksl_{(i_1+2)\cdots (i_2-1)}}.{i_2}},
&\qquad l=i_2\\
 \displaystyle
\sandmp{1}.{\Ksl_{2\cdots i_2}\Ksl_{(i_1+2)\cdots i_2}}.{(l+1)}\\
  \hskip 10mm
 \displaystyle\vspacer%
  \times
 \frac{%
       \sandmp{1}.{\Ksl_{2\cdots i_2}\Ksl_{(i_1+2)\cdots i_2}\Ksl_{l{}(l+1)}\Ksl_{(i_1+2)\cdots l}\Ksl_{(i_1+2)\cdots i_2}\Ksl_{2\cdots i_2}}.{1}}
      {\smash{\spa{l}.{(l+1)}}\vphantom{1}},
&\qquad (i_1+2)\leq l < i_2
\end{array}\right.
\\
%
&&\hspace*{-0.8cm} f_2(n;l,p,i_1,i_2)=\nonumber\\
&&\left\{\begin{array}{p{\widest}l}
$\displaystyle\vspacer%
\frac{\sandmp{i_2}.{\Ksl_{(i_1+2)\cdots i_2}\Ksl_{2\cdots (i_1+1)}}.{1}}
 {s_{(i_1+2)\cdots i_2}%
  \sandmp{1}.{\Ksl_{(i_2+1)\cdots n}\Ksl_{l\cdots i_2}\Ksl_{(i_1+2)\cdots (l-1)}\Ksl_{2\cdots (i_1+1)}}.{1}},$
&\qquad p=i_2 \\
$\displaystyle
\vspacer%
\frac{\spa{p}.{(p+1)}}
  {\sandmp{1}.{\Ksl_{(i_2+1)\cdots n}\Ksl_{(i_1+2)\cdots i_2}\Ksl_{(i_1+2)\cdots (l-1)}\Ksl_{l\cdots p}}.{(p+1)}},$
&\qquad l+1\leq p < i_2 \\
\end{array}\right.
\\
%
&&\hspace*{-0.8cm} f_3(n;l,p,i_1,i_2)=\nonumber\\
&&\left\{\begin{array}{p{\widest}l}
$  \displaystyle
\vspacer%
  \frac{\sandmp{1}.{\Ksl_{2\cdots (i_1+1)}\Ksl_{(i_1+2)\cdots i_2}}.{(i_2+1)}}
       {\sandmp{1}.{\Ksl_{(p+1)\cdots n}\Ksl_{(i_2+1)\cdots p}\Ksl_{(i_1+2)\cdots i_2}\Ksl_{2\cdots (i_1+1)}}.{1}},$
&\qquad l=i_2+1\\
$  \displaystyle
\vspacer%
 \frac{\spa{(l-1)}.{l}}{\sandmp{1}.{\Ksl_{(p+1)\cdots n}\Ksl_{l\cdots p}}.{(l-1)}},$
&\qquad l>i_2+1\\
\end{array}\right.
\end{eqnarray}
and~\cite{QPaper},
\begin{equation}
{\cal F}(l,p) = \sum_{i=l}^{p-1} \sum_{m=i+1}^{p} \ksl_i \ksl_m \,.
\label{Flpdef}
\end{equation}

In addition to spinorial collinear and multiparticle singularities, which
are genuine physical singularities of the amplitude, 
the various functions given above also contain spurious singularities.
These are of four kinds: (a) `planar' or `back-to-back' singularities
arising from the vanishing of spinor sandwiches like $\sand{1}.{\Ksl}.{j}$;
(b) `cubic' singularities from the vanishing of forms like
$\sandmp{1}.{\Ksl\Ksl'}.{j}$; (c) `open' and `closed quintic' singularities
from the vanishing of forms like $\sandmp{1}.{\Ksl_1\Ksl_2\Ksl_3\Ksl_4}.{j}$.
(Closed quintics correspond to $j=1$, and the sum of all momenta vanishing.)
While individual terms contain these singularities, they cancel in the
amplitude as a whole.  We have verified this cancellation numerically
through $n=12$. We have also verified the collinear and multiparticle
factorization numerically, and compared with fixed-order calculations of
the amplitude through $n=8$.


We thank Zvi Bern and Lance Dixon for extensive discussions and for providing
code for fixed-point amplitudes used as cross checks.
  We also thank Academic Technology Services at UCLA for
computer support.


\end{document}